**Epitaxial Growth of Anisotropic SnSe on GaAs(001) via Step-Edge Orientation Control**

Pooja D. Reddy[1], Zahra N. Heussen[2], Kunal Mukherjee[1*]

[1]Department of Materials Science and Engineering, Stanford University, Stanford, CA 94305, USA

[2]Department of Electrical Engineering, Stanford University, Stanford, CA 94305, USA

**Abstract**

Epitaxial growth of orthorhombic SnSe on cubic substrates is challenging due to lattice-symmetry mismatch and anisotropic bonding. Here we demonstrate that epitaxial films with sharp interfaces can be achieved for layered SnSe grown directly on on-axis and 4° miscut GaAs(001) substrates. The substrate miscut strongly influences the growth morphology, evolving from spirals on on-axis GaAs to a terraced structure on miscut GaAs. X-ray diffraction reveals that on-axis GaAs supports SnSe with two in-plane orientation variants, whereas the miscut substrate stabilizes a single orientation and introduces a small out-of-plane tilt. Accordingly, in-plane optical anisotropy is enhanced in the single variant film compared to the double variant, as determined by cross-polar reflectance. High-resolution TEM shows that the SnSe/GaAs interface is atomically abrupt and incoherent, characteristic of quasi–van der Waals epitaxy. We find a pronounced tendency for the zigzag edges of SnSe to align parallel to step edges on both substrates, and we show that step-skipping nucleation and layer growth on the miscut substrate leads to the additional tilt. These results establish direct SnSe/GaAs heteroepitaxy as a route to integrate anisotropic layered semiconductors with cubic platforms, and show that miscut substrates provide additional control over in-plane anisotropy.

---

[*] kunalm@stanford.edu

# I. INTRODUCTION

Anisotropic layered materials offer promising applications in optoelectronics due to their directionally dependent properties.[1–5] Tin selenide (SnSe), a layered orthorhombic semiconductor (*Pnma* space group, indirect 0.9 eV band gap),[6,7] has historically been studied in its bulk single crystal or polycrystalline form as a thermoelectric material, achieving the highest reported ZT values.[8] More recently, interest has expanded to microelectronic applications, including to SnSe's in-plane ferroelectricity, use in valleytronics, and role as a two-dimensional p-type channel material.[9–15] Prior work on SnSe and PbSnSe has also revealed an accessible cubic-to-orthorhombic phase transformation, suggesting potential for reconfigurable optical devices.[7,16–18] These applications require epitaxial thin films, typically on cubic substrates.

A central challenge in achieving single crystal epitaxial SnSe films is growing this orthorhombic crystal on the cubic substrates commonly used in industry. Additionally, SnSe exhibits layered, 2D bonding, making it important to understand how it grows on a three-dimensionally bonded substrate, a mechanism often described as quasi–van der Waals (vdW) epitaxy.[19] As an example, SnSe often nucleates in two orthogonal in-plane (IP) orientation variants on cubic substrates, producing 90° twin domains.[16,20] SnSe has directionally distinct refractive indices ($n$, $k$),[5,20,21] and so this twinning averages out macroscopic properties and introduces grain boundaries, that further reduce the functional anisotropy of the film and introduces scattering loss. Therefore, lattice constant mismatch, chemistry/bonding mismatch, and symmetry differences complicate epitaxial alignment and strain relaxation. Mortelmans et al. have addressed the symmetry issue by using an a-plane sapphire substrate with rectangular IP lattice symmetry.[20] We have previously addressed the chemistry mismatch with cubic PbSe buffer layers, but this retains two IP orientation variants due to the symmetry mismatch.[22]

A promising route that researchers have explored in other layered materials is to break the substrate surface symmetry to suppress twin formation. For instance, asymmetric substrate surface reconstructions in MBE can stabilize a single IP orientation variant of orthorhombic $Sb_2Se_3$.[23] Miscut substrates provide another route, where step edges act as preferential nucleation sites for 2D materials, enabling oriented grain growth in $WSe_2$ and promoting spiral growth in SnS.[24–26] In $Bi_2Se_3$, growth on miscut sapphire reduces twin density, although this selectivity diminishes as the film thickens and layers overgrow the steps.[27] These studies highlight the importance of step-mediated nucleation for controlling orientation in low-symmetry layered semiconductors. This motivates a detailed understanding of the bonding between SnSe and cubic substrates and raises the key question of whether epitaxial SnSe films can be oriented on cubic substrates to preserve their IP birefringence and avoid twin formation. In this work, we investigate the direct growth of SnSe on GaAs(001) and its interfacial structure and epitaxial quality. We show that orthorhombic SnSe can be grown directly on cubic GaAs(001) with atomically sharp interfaces through quasi–vdW epitaxy, and that the use of miscut GaAs substrates enables step-mediated, orientation-selective bonding that suppresses twinning and stabilizes a single IP orientation.

## II. METHODS

SnSe thin films were grown on a Riber Compact 21 IV–VI system using a compound source SnSe dual effusion cell. Substrates used were As-capped homoepitaxial GaAs(001) prepared in a Veeco Gen III chamber ex situ, and transferred into the IV-VI chamber for growth. The substrate was heated to 420 °C to thermally desorb the amorphous As cap exposing a high quality GaAs surface. Previous work has shown that a PbSe surface treatment can seed epitaxial layered SnSe on cubic GaAs. A PbSe flux was supplied to the substrate at 420 °C for 30 s to modify the surface chemistry without forming a PbSe layer.[16] The substrate was then cooled to 300 °C and SnSe was grown at a rate of approximately 0.3 Å/sec.

A 25 nm on-axis SnSe/GaAs(001) film was characterized using X-ray diffraction (XRD), atomic force microscopy (AFM), and transmission electron microscopy (TEM). Polarized optical microscopy was performed on approximately 70 nm thick on-axis and 4° miscut SnSe/GaAs(001) films to more directly compare reflectance. The 70 nm miscut film was additionally characterized using XRD, AFM, and TEM. AFM measurements were performed using a Park Systems NX-10 microscope. XRD measurements, including reciprocal space maps (RSMs) and symmetric 2θ–ω scans were collected on a PANalytical Empyrean and X'Pert Pro respectively, using Cu Kα1 radiation. RSMs were acquired in grazing exit geometry, and symmetric 2θ–ω scans were collected using an open detector configuration. Scanning transmission electron microscopy high-angle annular dark-field (STEM HAADF) images were collected on a Thermo Fisher Spectra 300 operated at 300 kV. Polarized optical microscopy measurements were performed using a Nikon LV100 microscope with a polarizer and analyzer.

## III. RESULTS AND DISCUSSION

### A. Surface morphology and step structure

Since the substrate miscut can break the surface symmetry and promote preferential IP orientations,[24,25,27] we grew SnSe films on both on-axis GaAs(001) and 4° miscut toward the [111]B direction GaAs(001) substrates by molecular beam epitaxy (MBE) at ~300°C using a brief PbSe-dose surface preparation. Figure 1 compares surface morphology using AFM scans and corresponding line profiles for the 25 nm on-axis and 70 nm miscut samples. In the miscut sample, the step edges run horizontally across the scan, with the vertical scan direction corresponding to descending steps. The surface morphology of the on-axis sample is consistent with previous reports of SnSe and similar layered systems, where screw dislocations mediate growth.[25,28,29] This film is grown directly on high-quality GaAs, so rather than growing via inherited dislocations from a crystalline substrate or buffer layer,[22,30] the observed spirals in Figure 1ai likely originate from dislocations related to island coalescence and local pinning of the growth front at domain boundaries or substrate step edges due to the weak OP bonding in 2D materials.[31–33] Terrace widths of SnSe on on-axis GaAs range from 35 to 50 nm and step edges are predominantly $\frac{a}{2}$ in height (Fig. 1aii), consistent with the bilayer stacking of SnSe separated by vdW bonds. This is consistent with SnSe and related materials often having $\frac{a}{2}$ screw dislocations.[22,25] Occasional $\frac{a}{4}$ or $\frac{3a}{4}$ steps appear in our AFM traces when the scan crosses incomplete or overlapping spiral turns, rather than representing true step-height variations. Trace 6 in Figure 1aii illustrates this behavior, as the scan crosses over the top of a spiral mound. Genuine $\frac{a}{4}$ or $\frac{3a}{4}$ steps would require breaking stronger intra-bilayer bonds and would produce complex stacking faults, which are not observed in TEM data.

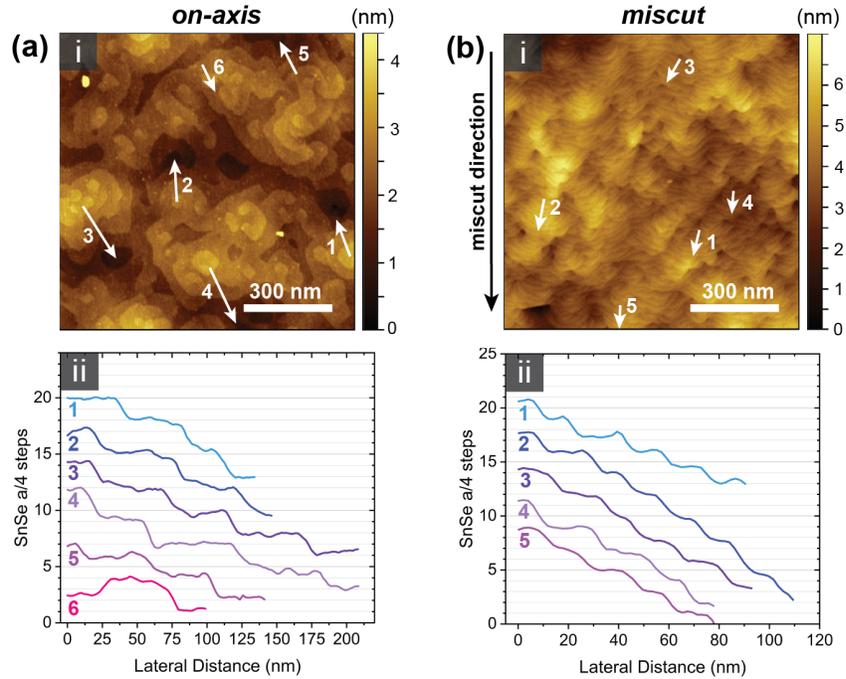

**Figure 1.** AFM micrographs and corresponding line scans of (a) SnSe on on-axis GaAs and (b) SnSe on 4° miscut GaAs with miscut direction marked. (i) AFM area scans, with the positions of the line scans indicated, and (ii) line profiles, plotted in units of a/4 SnSe step height. The SnSe film on on-axis GaAs shows the typical spiral growth mode, whereas the film on the miscut substrate shows step-flow growth along the miscut direction, with narrower terrace widths. In both cases, step heights are predominantly a/2 SnSe.

The SnSe grown on a miscut GaAs substrate has a distinctly different surface morphology. Rather than spiral growth islands seen for the on-axis film, the miscut film surface retains a stepped structure along the substrate miscut direction (Fig. 1ai). The step edges are consistently a/2, as is expected for SnSe, with narrower terrace widths close to 20 nm. The step edges are pinned at discrete sites which are often also surface depressions. These pinning points are likely screw dislocations, which have been shown to cause similar surface depressions and pinning in step edge flow growth.[34–36] The step edges bow outward between these pinning points, giving the steps a puckered appearance. Before a complete spiral rotation can form around a screw dislocation, the next advancing step front spiral growth overtakes it, producing the continuous step-flow morphology.

## B. Crystallographic orientation and in-plane variant selection

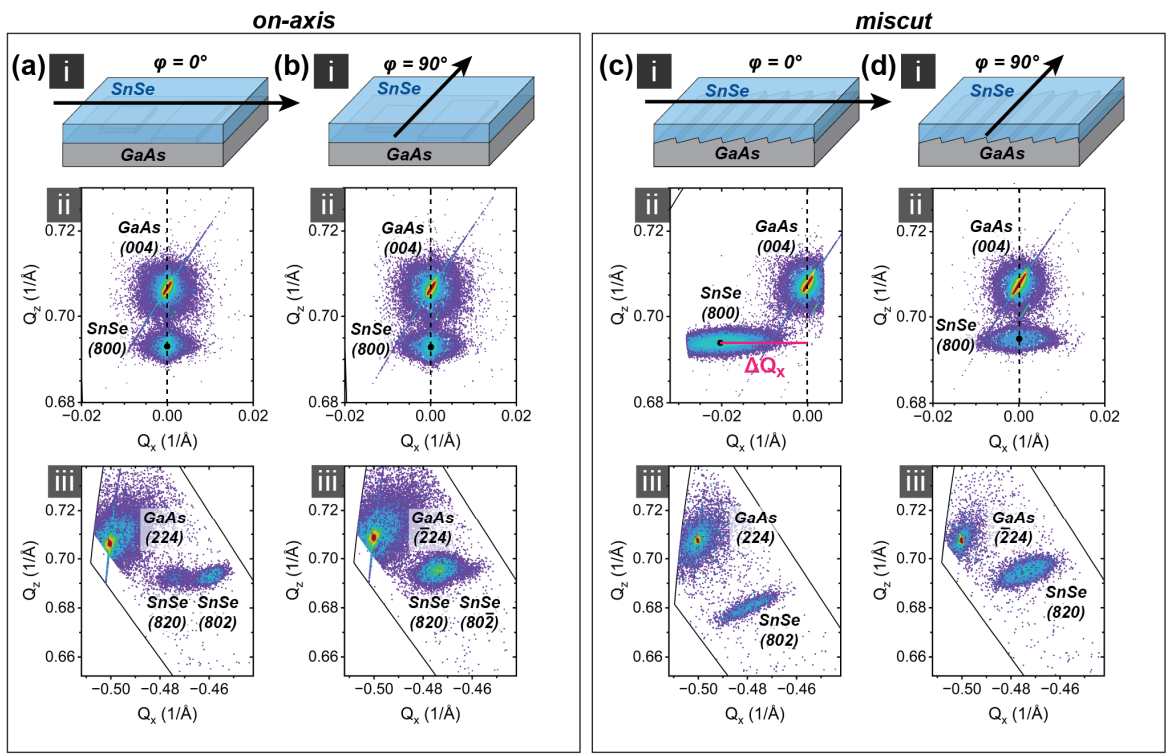

**Figure 2.** RSMs of SnSe films on on-axis and 4° miscut GaAs(001). For each sample, (i) shows a schematic of the measurement geometry with the incident X-ray beam azimuth direction marked, (ii) the symmetric RSM around GaAs (004) with the $Q_x = 0$ line marked, and (iii) the asymmetric RSM around GaAs (224) with film peaks labeled. (a, b) SnSe on on-axis GaAs at azimuths $\varphi = 0°$ and 90°, respectively. (c, d) SnSe on 4° miscut GaAs toward the [111]B direction at $\varphi = 0°$ (perpendicular to step edges) and 90° (parallel to step edges) respectively. For the on-axis films, no $Q_x$ offset is observed in the symmetric scans, indicating the films are untilted, and two film peaks appear in the asymmetric RSMs, corresponding to two in-plane orientations. For the miscut case, a $Q_x$ offset is observed in the $\varphi = 0°$ azimuth, consistent with a film tilt, and only one film peak is present in the asymmetric RSMs, indicating a single in-plane orientation.

Figure 2a presents XRD characterization of a 25 nm SnSe film grown on nominally on-axis GaAs(001), measured with the sample rotated to $\varphi = 0°$ and $\varphi = 90°$ (relative to GaAs $\langle 110 \rangle$) (Fig. 2a and 2b). RSMs around the symmetric GaAs (004) reflection confirm an epitaxial relationship between film and substrate, with the SnSe [100] direction parallel to GaAs [001] (Fig. 2aii and 2bii). Symmetric 2θ-ω scans confirm that the SnSe film has only one out-of-plane (OP) orientation on GaAs (Fig. S2). The lack of any $Q_x$ offset between the GaAs (004) and SnSe (800) peaks indicates no relative film tilt for SnSe on on-axis GaAs(001).

RSMs around the asymmetric GaAs (224) reflection for the 25 nm SnSe on on-axis GaAs(001) sample resolves the SnSe (820) and (802) peaks (Fig. 2aiii, 2biii). The presence of both film peaks indicates that the orthorhombic unit cell, with its rectangular base, adopts two orthogonal orientations on the cubic GaAs substrate, creating IP-oriented variants. Specifically, the IP lattice directions of SnSe align along orthogonal GaAs $\langle 110 \rangle$ directions to minimize lattice mismatch, consistent with previous reports.[16,22] From the asymmetric RSMs, we extract orthorhombic lattice constants of a =11.51 Å, b = 4.22 Å, and c = 4.35 Å for the SnSe thin film on on-axis GaAs(001), differing somewhat from bulk values (a = 11.5 Å, b = 4.15 Å, c = 4.45 Å).

The brief PbSe-dose surface preparation step used for our MBE growth process results in a Se-stabilized 2×1 asymmetric surface reconstruction and may be similar to 2×1 Se-treated GaAs surfaces reported in the literature.[16,37] While asymmetric surface reconstructions can preferentially orient some vdW bonded materials, those on GaAs are not sufficient to stabilize a single IP orientation of SnSe. When the sample is measured at φ = 0° and φ = 90°, the signal from observed peaks interchange: (802) appears as (820) and vice versa. IP orientation variant fractions were quantified by integrating peak intensities with structure-factor corrections (S1), yielding ~20% of the film volume in one orientation and ~80% in the other. More generally, we find that the SnSe nucleation process is highly sensitive to the GaAs surface preparation, with changes in the GaAs oxide-desorption step, the PbSe dose duration, or temperature variability shifting the orientation balance between 95:5 and 65:35, even for nominally identical growth conditions. Because this outcome is difficult to reproducibly control, an alternative approach is needed to more robustly bias orientation for applications where the full IP anisotropy is required.

An approach to preferential IP orientation is breaking the substrate surface symmetry using a miscut wafer.[24,25,27] We used GaAs substrates with a 4° miscut toward the [111]B direction, which produces a vicinal surface with regular step-edge arrays along the ⟨110⟩ IP directions. If SnSe maintains the same epitaxial relationship as on on-axis substrates, with [100] oriented OP and IP axes aligned along GaAs ⟨110⟩, then the SnSe armchair and zigzag IP directions should lie parallel and perpendicular to the step edges. Consistent with this expectation, symmetric 2θ–ω scans show that a 70 nm SnSe film grown on the miscut substrate retains the dominant SnSe [100] OP orientation (Fig. S2). GaAs (004) RSMs along φ = 0° and 90° also reveal that the (800) peak of the film is noticeably tilted relative to the substrate. We quantify this tilt from the $Q_x$ offset along φ = 0° as tilt angle = arctan($\Delta Q_x/Q_z$). This analysis shows that the film develops an additional 1.6° tilt in the direction perpendicular to the step edges (Fig. 2cii), while no tilt is observed parallel to the step edges (Fig. 2dii). The tilt mosaicity, that is spread in tilt values, is also significantly greater for the SnSe film on the offcut substrate than that of the on-axis case.

The IP structure was examined using RSMs around the asymmetric (224) reflection of GaAs, acquired with the incident X-ray beam azimuths perpendicular (φ = 0°, Fig. 2c) and parallel (φ = 90°, Fig. 2d) to the step edges. These reveal a marked difference from films grown on on-axis substrates. The film on the miscut substrate only has one film reflection at each azimuth. After correcting for the measured film tilt (Section S2), the extracted lattice constants are a = 11.52 Å, b = 4.22 Å, and c = 4.35 Å, essentially identical to that on the on-axis substrate. Thus, the miscut influences orientation selection and tilt without altering SnSe lattice parameters significantly. The calculated IP lattice constants are distinct (b ≠ c), as expected for the orthorhombic SnSe structure and allow us to assign them to specific crystallographic planes. At φ = 0°, only the (802) reflection associated with the IP *c* (armchair) direction is observed (Fig. 2ciii), while at φ = 90° only the (820) peak associated with the IP *b* (zigzag) direction appears (Fig. 2diii). Given these distinct IP lattice constants, the presence of only one film peak at each azimuthal orientation is striking evidence for a single IP SnSe orientation variant. Specifically, we find step edge mediated selection of the zigzag direction parallel to the GaAs step edge (and armchair direction perpendicular to the step edge). This selective bonding is independently confirmed using TEM. We discuss later how the zigzag edge geometry may provide a more favorable step-edge interaction.

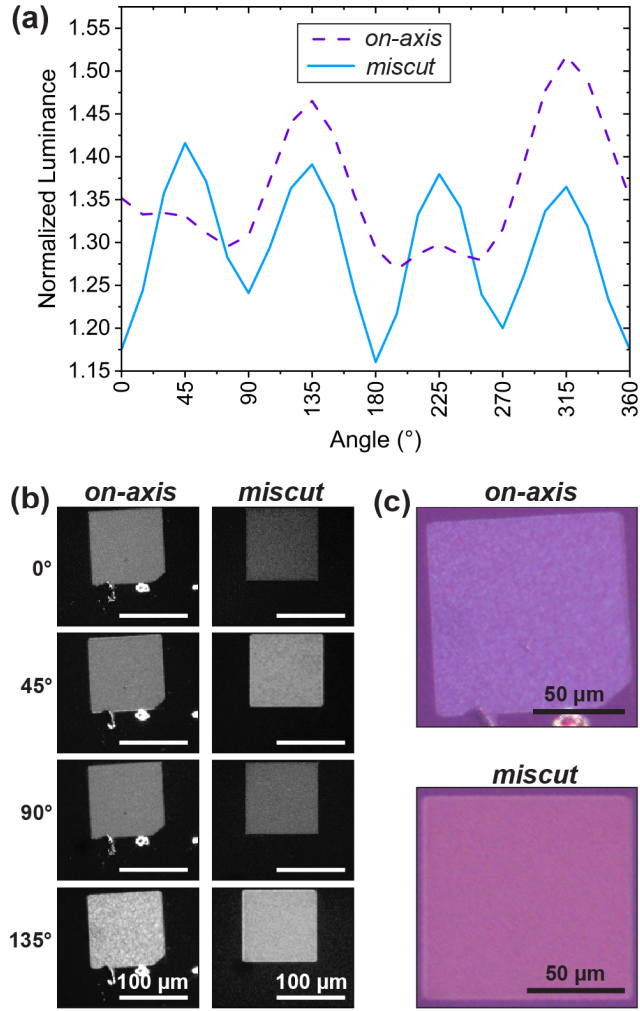

**Figure 3.** Polarized-optical (crossed-polarizer) reflection data from SnSe mesas on on-axis GaAs(001) and 4° miscut GaAs(001). (a) Film luminance normalized by background GaAs luminance, as a function of sample rotation angle, for the SnSe films, where the 0° orientation is defined as the GaAs <110> direction. (b) Grayscale luminance micrographs at 0° and 90° (dark azimuths), and 45° and 135° (bright azimuths) for both films. (c) RGB micrographs at 135°, a bright azimuth for SnSe on on-axis and miscut GaAs(001), highlighting differences in intensity and hue variation.

### C. Spatially resolved optical anisotropy

We use cross-polarized microscopy in reflection geometry to assess how differences in the IP orientation variants in the on-axis and miscut SnSe films lead to measurable changes in optical properties and their spatial uniformity. Mesas were etched into two ~70 nm SnSe films on on-axis and 4° miscut GaAs substrates. XRD indicated that the on-axis film contains two IP orientation variants in a 75:25 ratio, whereas the miscut film is single-variant. In this reflection-mode configuration, linearly polarized light is incident on the film surface, and the reflected signal is analyzed between crossed polarizers as the sample is rotated.

Both SnSe films display a four-fold modulation in normalized luminance (Fig. 3a), consistent with in-plane optical anisotropy. Grayscale images of the two mesas at select rotation angles are shown in Figure 3b. This behavior is expected because SnSe is both birefringent ($n_b \neq n_c$) and dichroic ($k_b \neq k_c$) at visible wavelengths. The single IP orientation film on the miscut substrate exhibits a much clearer four-fold brightness variation between minima and maxima. For an orthorhombic film with a single IP orientation, the reflected intensity has minima along the projected optic axes (dark azimuth) and maxima approximately 45° away (bright azimuth).[20,21,38–40] The on-axis sample has a much weaker four-fold signature as expected from the two IP orientation variants. An additional twofold brightness variation, which is not expected from crystallography, is also present and likely arises from grain boundaries, or optical bias and misalignment in the microscopy setup.[20]

Higher magnification RGB images acquired at 135°, a bright azimuth, reveal spatial texture in both films (Fig. 3c). In the miscut film, which contains a single IP orientation, the observed intensity and hue variations are likely dominated by surface texture and small angle grain boundaries. In contrast, the on-axis film exhibits a patchier appearance with larger variations in intensity and hue, consistent with polarization-dependent contrast from its domain structure, in which micron-scale grains of different IP orientation variants are preferentially highlighted under crossed polarizers. Overall, while XRD provides the averaged orientation distribution, cross-polarized optical microscopy offers a complementary probe of optical anisotropy and uniformity achieved on GaAs, on the tens-of-microns length scale.

**D. Interface structure and orientation mechanism**

High-resolution TEM provides insight at an even finer length scale, revealing how layered SnSe bonds to the 3D-bonded covalent GaAs and why a miscut substrate successfully stabilizes one IP orientation variant of SnSe.

*Double orientation variant SnSe on on-axis GaAs*

Figure 4 shows atomic resolution TEM images of the GaAs/SnSe interface on on-axis GaAs atomically flat terraces. Figures 4aii and 4bii are excerpts of atomic resolution images of the interface for domains with zigzag and armchair directions parallel to the incident electron beam, respectively. The GaAs substrate is nominally selenium-terminated as expected from the surface treatment of this GaAs(001) in MBE and corroborated using STEM HAADF contrast (Fig. S3).[37,41,42] In the armchair-oriented variant (Fig. 4aii), the crystal structure of SnSe and GaAs along GaAs⟨011⟩ coincide at distance of 11 SnSe/12 GaAs units. For the zigzag-oriented variant (Fig. 4bii), a similar relationship emerges with alignment every 17 SnSe/18 GaAs units. This 'n/n+1' lattice correspondence is commonly seen in domain-matching epitaxy.[43] The weak interfacial bonding between the layered SnSe and GaAs also enables epitaxial growth without traditional misfit dislocations defects at the interface, known as quasi-vdW epitaxy.[44,45] In this regime, large lattice mismatches can be tolerated because strict bond alignment between the film and substrate is unnecessary,[45,46] as the dangling bonds of the 3D-bonded substrate are passivated/terminated before adding the 2D-bonded material on top.[19,47–49] Previous studies have shown that GaAs(001) dangling bonds can be passivated under Se flux, converting the As-terminated (2×4) surface to a Se-stabilized (2×1) reconstruction via As–Se substitution.[37,50–54] Related work has also demonstrated that layered selenides such as GaSe and $In_2Se_3$ can be grown on GaAs and InP with interfacial passivation, either from Se species or binary selenides bonding that saturate surface bonds and form a passivating layer.[48,55–57] Although we lack atomic-resolution data to directly determine the interfacial chemistry or whether passivation arises from Se or PbSe species, we speculate that the PbSe dose step plays a comparable role, passivating the GaAs surface and enabling quasi-vdW epitaxy of SnSe.

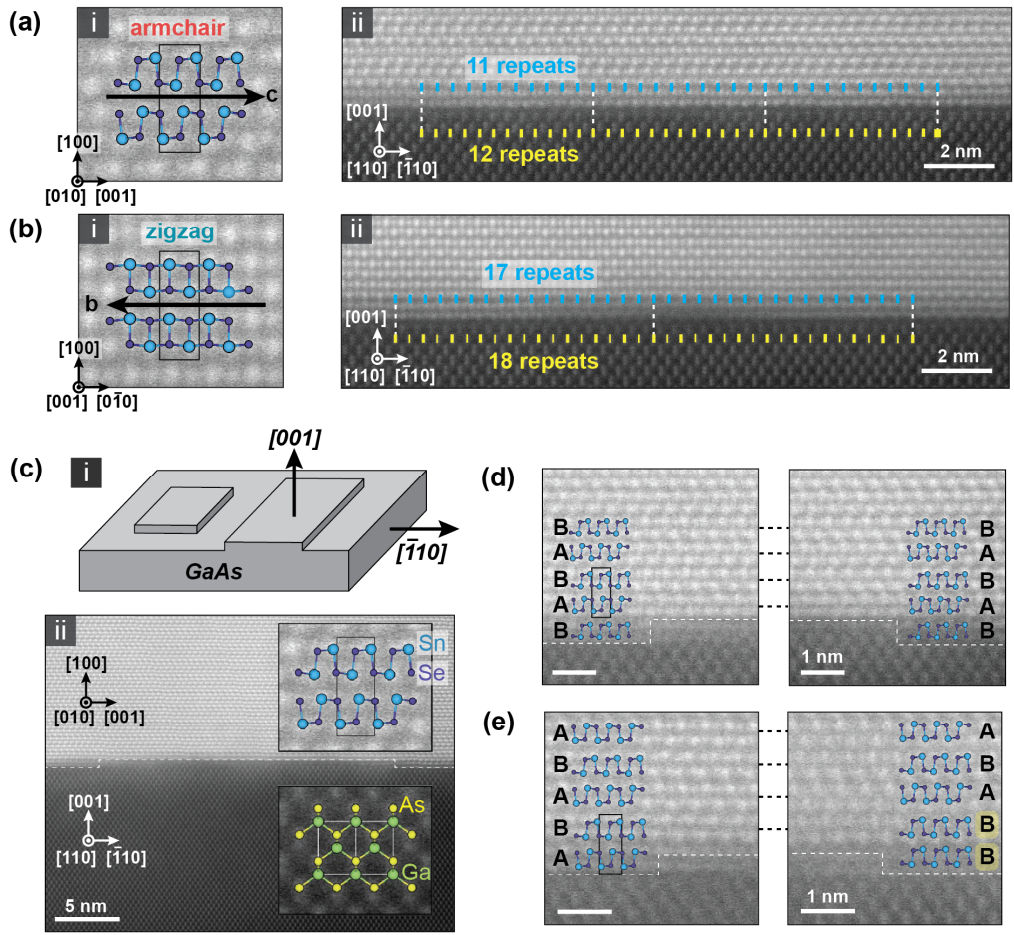

**Figure 4.** STEM micrographs of SnSe on on-axis GaAs(001). (a) Armchair-oriented SnSe grain on GaAs, and (b) zigzag-oriented grain. (i) Close-up views highlighting the IP orientation, and (ii) corresponding micrographs of the interface with overlaid markings emphasizing the OP alignment. Lattice directions are indicated in each panel. (c) (i) Schematic and (ii) STEM micrograph of an SnSe island on the GaAs surface with lattice directions marked, showing step edges that serve as IP bonding sites between SnSe and GaAs. Only armchair-oriented grains are observed to bond laterally with GaAs step edges via their zigzag edges. (d, e) Step edges on either side of a GaAs island, where both A and B layers of SnSe bond to step edges on opposing sides. Occasional AA or BB stacking is observed, which self-corrects to the normal AB sequence with continued layer growth.

While the OP bonding at the interface lacks registry due to vdW bonds, we expect IP covalent bonding between SnSe and on-axis GaAs at terrace step edges (Fig. 4ci). On the on-axis substrate, the terrace steps we see are double step edges $\frac{c}{2}$ (two atomic layers). Atomic-resolution STEM HAADF images of a small terrace (Fig. 4cii) and representative step edges (Fig. 4d, 4e) on the substrate reveal that the zigzag edge of SnSe consistently aligns parallel to GaAs step edges—that is the armchair orientation is perpendicular to the step edge—and indeed the SnSe bilayers are bonded to GaAs. This zigzag edge preference and bonding was seen at all ~5 observed step edges of the armchair-oriented grains. No step edges or bonding to GaAs was noted, however, for the zigzag oriented grains suggesting that either the incidence of their armchair edges lying parallel to the step edges are low or these domains have zigzag edges bonding to steps that are running parallel to the TEM foil (and hence cannot be seen).

We note two features of the zigzag edge bonding. First, both left- and right-oriented SnSe bilayers (labelled as A or B) are seen bonding to the step edge (Fig. 4d, 4e). Second, the GaAs double step height closely matches the thickness of a SnSe bilayer $\frac{a}{2}$ (nominally 5.65 Å vs 5.75 Å), which promotes bilayer nucleation at GaAs these double steps. As a result, separate nucleation events can produce AA-stacked metastable domains near the interface upon coalescence. Yet, the film subsequently reconfigures into the stable AB stacking beyond the first few layers (Fig. 4e). The flexibility of SnSe bonding is further evident within individual terraces, where the bilayer stacking can locally switch type (A→B or B→A, Fig. S5). We have observed this interfacial AA stacking in SnSe previously with random nucleation orientation and step-flow overgrowth,[22] and given the reported ferroelectricity of AA-stacked (or β' phase) SnSe,[9,12,13] merits further study to explore how AA stacking might be stabilized in thin films beyond a few bilayers.

We propose that (Sn)–(Se) bonds at Se-terminated GaAs step edges likely provide favorable bonding for the SnSe zigzag orientation. Zigzag edges of SnSe nominally have either Sn dangling bonds or Se dangling bonds presented in a uniform 1D motif.[58] Armchair edges, on the other hand, have mixed chemistry and present nonequivalent edge sites associated with the puckered bilayer. Additionally, the zigzag bond spacing in the unit cell is nominally lower lattice-mismatched (5.6%) vs. the armchair edge (8.8%) to the <110> bond spacing of GaAs. Together, these factors suggest that Sn-terminated zigzag edges can couple more strongly to Se-terminated GaAs step edges. Nevertheless, a specific IP bonding motif is difficult to discern at this point as the interface at step appears blurred (Fig. 4c-e), likely due to stacking disorder and meandering island facets that overlap GaAs and SnSe lattices in projection. Even so, the SnSe atomic layers appear to align with anion atom positions in the GaAs structure (Fig. 4d, 4e).

Although we do not image the orthogonal azimuth of GaAs, we can infer that the two unequal fractions of IP-orientation variants seen in XRD arise from this preference of the zigzag edge to orient parallel to both [110] and [$\bar{1}$10] step edges. Step edges on Se-terminated GaAs(001) differ chemically and may have different reactivities,[59,60] so the persistence of both orientations suggests that zigzag edge SnSe bilayers can align parallel to multiple step-edge chemistries. The unequal fractions seen for on-axis growths likely reflect asymmetric terrace shapes on the substrate rather than armchair edges aligning to step edges.

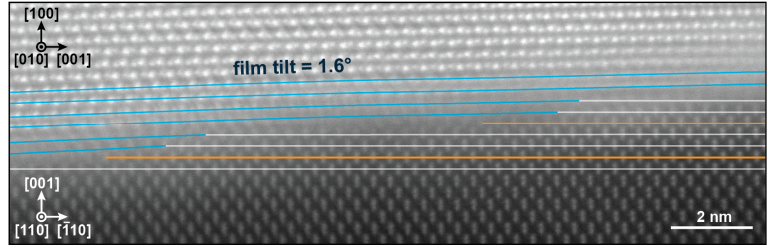

**Figure 5:** Atomic-resolution STEM micrographs of SnSe grown on GaAs(001) 4° miscut perpendicular to the miscut and step edge direction, with overlaid markings highlighting IP alignment. Lattice directions of SnSe and GaAs are marked in. The SnSe film is tilted with respect to the GaAs lattice. SnSe atomic layers (blue lines) connect to some GaAs atomic layers (white lines), while orange lines mark GaAs layers that remain unbonded.

*Single orientation variant SnSe on miscut GaAs*

The preference for the SnSe zigzag direction to align parallel to GaAs step edges is consistent with the epitaxial relationship inferred from XRD of SnSe on the 4° miscut GaAs substrate. Single IP orientation is obtained by simply eliminating substrate step edges in the orthogonal direction. However, the SnSe grown on the miscut GaAs develops an additional tilt relative to the substrate. To examine this in more detail, we prepare a cross-sectional TEM specimen perpendicular to the step edge direction to capture the interface and epitaxial tilt. For clarity, the micrograph in Figure 5 has been rotated by the substrate miscut angle so that the GaAs lattice directions appear horizontal and vertical. Only armchair-oriented grains are observed in the miscut film, contrasting with the two-orientation growth on on-axis substrates.

At the interface, we see the same domain-matching relationship seen on the on-axis sample along the armchair direction. The OP bonding remains incoherent, maintaining the vdW-like bonding character observed on the on-axis substrate. However, the IP bonding shows additional nuance. A detailed look at the atomic planes (Fig. 5) reveals that SnSe layers do not bond to every available GaAs atomic step (marked in orange), typically these skipped steps are single steps ($\frac{c}{4}$). The SnSe bilayer from the preceding step nevertheless conforms to the step morphology by "drooping" over the skipped step edges. This conformal sagging accounts for the ~1.6° average film tilt measured by XRD.

The measured tilt can be explained by a nucleation and step-skipping mechanism. As mentioned previously, a bilayer of SnSe (nominally 5.75 Å tall) is heavily mismatched to GaAs single steps (2.82 Å), and 1.8% taller than even a GaAs double step (5.65 Å). If SnSe bilayers are nucleated at every step edge (irrespective of single or double steps), then the SnSe layers have to distort upwards and further increase the tilt of the planes compared to that of the substrate, a mode previously termed as step-climbing epitaxy in $Bi_2O_2Se$ growth on $SrTiO_3$ but previously noted also on $WSe_2$ growth on miscut sapphire.[24,61] On the other hand, if the SnSe nuclei density is low, or suppressed at single steps due to mismatch, then the bilayers from preceding step edges would conformally drape across the skipped steps.[24] For extremely low density of nuclei, the film would largely counter the offcut of the substrate. For intermediate densities, the film would partially counter the offcut as in the case of our samples. Thus, it seems the net epilayer tilt may potentially be tunable via growth conditions and is a subject of future work.

**IV. CONCLUSIONS**

We successfully synthesized epitaxial films of orthorhombic SnSe directly on cubic GaAs(001) substrates, both on-axis and with a 4° miscut, with atomically sharp interfaces. SnSe on on-axis GaAs has a spiral growth mode, and two IP orientation variants, which indicates that SnSe can bond to chemically distinct step edges. On both substrates, high-resolution TEM shows that the SnSe/GaAs interface lacks coherent OP registry, characteristic of quasi–vdW epitaxy between a 2D-bonded film and a 3D-bonded substrate. The interface remains sharp and free of misfit dislocations, and the PbSe surface dose likely passivates GaAs dangling bonds to enable defect-free nucleation. We additionally observe that the SnSe zigzag edge selectively aligns parallel to GaAs step edges. Taking advantage of this selective IP bonding, SnSe grown on a 4° miscut GaAs substrate exhibits a uniform step-flow morphology, a film tilt relative to the GaAs, and critically, eliminates the twin variant to stabilize a single IP orientation of orthorhombic SnSe on a cubic substrate. The selective IP bonding together with weak OP coupling allows SnSe layers to conform to the terrace morphology, occasionally skipping substrate atomic planes, producing a film tilt. Polarized optical microscopy of the single-oriented film exhibits strong fourfold brightness modulation, reflecting IP optical anisotropy and linking structural anisotropy to optical response.

 Overall, these results establish that direct SnSe/GaAs heteroepitaxy yields sharp interfaces through quasi–vdW coupling. Breaking substrate symmetry via miscut provides a practical means to control IP film orientation, through step-mediated, orientation-selective interactions. Our work offers a model for integrating anisotropic layered semiconductors with conventional cubic platforms, which is an essential step toward scalable, orientation-engineered optoelectronic devices exploiting SnSe's intrinsic optical anisotropy. Because this study focused on a single miscut magnitude and azimuth, future work that systematically varies vicinal geometry would be valuable. In particular, such experiments could elucidate how unit-cell mismatch and finite step density compete during growth to set the resulting film orientation and tilt, an increasingly relevant question as vdW materials are pursued in thin film form rather than as exfoliated flakes. More broadly, this work suggests that miscut surfaces can serve as a general geometric control knob for IP orientation and average film tilt in low-symmetry quasi-vdW heteroepitaxy beyond the specific SnSe/GaAs system studied here.

## SUPPLEMENTARY MATERIAL

See the supplementary material for details on tilt and IP variant fraction calculations, as well as additional XRD data and TEM micrographs.

## ACKNOWLEDGMENTS

This work was funded by the NSF CAREER award under Grant No. DMR-2036520. We made use of the Stanford Nano Shared Facilities (SNSF) supported by the NSF under Award No. ECCS-2026822 for various materials characterization techniques. P.R. gratefully acknowledges support from the National Science Foundation Graduate Research Fellowship under Grant No. DGE-1656518, as well as the Stanford Graduate Fellowship in Science and Engineering, and the P.D. Soros Fellowship for New Americans.

## AUTHOR DECLARATIONS

Conflict of Interest
The authors have no conflicts to disclose.
Author Contributions
Pooja D. Reddy: Formal analysis (equal); Investigation (lead); Methodology (lead); Validation (lead); Writing – original draft (equal). Zahra Heussen: Investigation (supporting); Writing – review & editing (equal). Kunal Mukherjee: Conceptualization (equal); Formal analysis (supporting); Funding acquisition (lead); Methodology (supporting); Validation (supporting); Writing – original draft (equal).

## DATA AVAILABILITY

The data that support the findings of this study are available from the corresponding author upon reasonable request.

[61] X. Zhou, Y. Liang, H. Fu, R. Zhu, J. Wang, X. Cong, C. Tan, C. Zhang, Y. Zhang, Y. Wang, Q. Xu, P. Gao, and H. Peng, "Step-Climbing Epitaxy of Layered Materials with Giant Out-of-Plane Lattice Mismatch," Advanced Materials **34**(42), 2202754 (2022).

SUPPLEMENTARY MATERIAL

**Epitaxial Growth of Anisotropic SnSe on GaAs(001) via Step-Edge Orientation Control**

Pooja D. Reddy[1], Zahra N. Heussen[2], Kunal Mukherjee[1]

[1]Department of Materials Science and Engineering, Stanford University, Stanford, CA 94305, USA

[2]Department of Electrical Engineering, Stanford University, Stanford, CA 94305, USA

**S1: Orientation Calculation for *Pnma* SnSe**

SnSe has two IP orientations when grown on cubic substrates. The ratio of these two IP orientations can change, depending on surface preparation steps and if any other changes are made to break the symmetry of the cubic substrate surface, such as using an offcut substrate. It is possible to estimate the relative fraction of the two IP orientation variants from the integrated intensities of the (802) and (820) reflections, which correspond to grain populations of the two variants. This is because this comparison involves reflections from the same phase, so experimental factors that influence XRD intensity such as geometric corrections, absorption, and Lorentz–polarization effects, are essentially identical for both the (802) and (820) peaks and cancel out. Therefore these calculations can be considered quantitative.

Peak intensities are calculated from RSMs by subtracting a background and counting the intensity within an ellipse that defines the boundaries of the SnSe (802) and (820) peaks. These values are then normalized by the GaAs substrate peak intensity to get normalized integrated intensity values. Structure factor values came from XRD diffraction patterns simulations of SnSe using the CrystalMaker software package: $|F_{Pnma,802}|=97$, $|F_{Pnma,820}|=164$, where intensity is proportional to $|F|^2$. We used the below equations for calculations.

$$fraction_{(802)} = \frac{\frac{I_{Pnma,802}}{|F_{Pnma,802}|^2}}{\frac{I_{Pnma,802}}{|F_{Pnma,802}|^2} + \frac{I_{Pnma,820}}{|F_{Pnma,820}|^2}}$$

$$fraction_{(820)} = \frac{\frac{I_{Pnma,820}}{|F_{Pnma,820}|^2}}{\frac{I_{Pnma,802}}{|F_{Pnma,802}|^2} + \frac{I_{Pnma,820}}{|F_{Pnma,820}|^2}}$$

$$fraction_{(820)} + fraction_{(802)} = 1$$

These fractions can be corroborated by performing calculations on asymmetric RSMs at both the $\varphi = 0°$ and $\varphi = 90°$ azimuths. In this geometry, the variant that appears as the (802) reflection at $\varphi = 0°$ will correspond to the (820) reflection at $\varphi = 90°$, and vice versa. The extracted orientation fractions should therefore invert between the two azimuths, providing a useful internal consistency check for the calculation.

## S2: Calculating and Accounting for Film Tilt

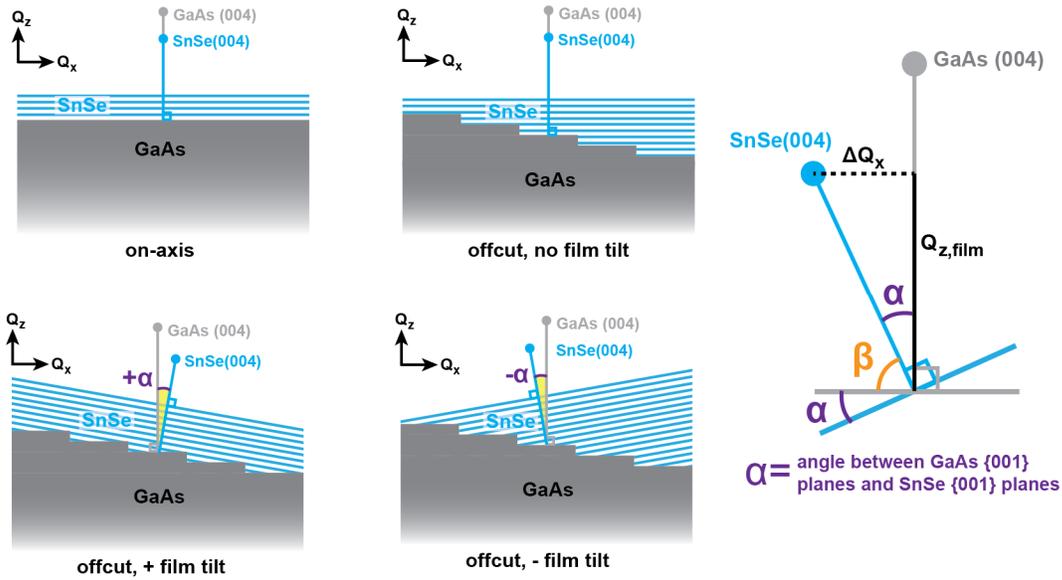

**Figure S1:** Schematics of the reciprocal space vectors for the (004) planes of GaAs and SnSe in $Q_x$-$Q_z$ space, visualized as normal to the corresponding lattice planes for various cases of SnSe on GaAs including, on-axis substrate, an offcut substrate with no film tilt, and an offcut substrate with film tilt in either direction. The enlarged schematic on the right hand side illustrates how the relative tilt between the substrate and film can be calculated, based on the position of the OP reflections from a symmetric RSM.

When measuring film tilt using OP reflections from a symmetric RSM, it manifests as a rotation about the $Q_y$ axis when the map is collected $\varphi = 0°$ and/or $\varphi = 90°$, with fixed tilt ($\chi$) (usually set at zero). We set these azimuths in the SnSe/GaAs measurements, along the step edge direction of the GaAs substrate, or $\langle 110 \rangle$ directions, and the IP directions of SnSe. Although tilt is a three-dimensional vector quantity, in this geometry it has a zero component along $Q_y$. If the film is tilted only with respect to the GaAs step edges (i.e., along one azimuth), the rotation about $Q_y$ will appear in only one of these orientations.

Figure S1 shows schematics of the reciprocal space vectors for the (004) planes of GaAs and SnSe (as would be captured by a symmetric RSM) for different cases, including, on-axis substrate, an offcut substrate with no film tilt, and an offcut substrate with film tilt in either direction. The reciprocal space vectors can be visualized as normal to the corresponding lattice planes. Therefore, when the film is tilted relative to the substrate, the film's reciprocal-space vector rotates, resulting in a shift of the film peak in the RSM. The tilt angle of a thin film with respect to the substrate can be expressed as follows.

$$\alpha = \tan^{-1}\left[\frac{\Delta Q_x}{Q_{z,film}}\right] = \tan^{-1}\left[\frac{Q_{x,film} - Q_{x,substrate}}{Q_{z,film}}\right]$$

If an RSM is collected along an azimuth where the film is tilted, the lattice constants calculated directly from the measured $Q_x$ and $Q_z$ values will be inaccurate, because the standard equations used to convert between reciprocal-space coordinates ($Q_x$, $Q_z$) and real-space lattice parameters assume no film tilt.

In the following analysis, we assume that the film tilt occurs along a single azimuth and is zero along the orthogonal direction. If the film is tilted along both azimuths, a more complex 3D correction is required. When scanning around the GaAs (224), the SnSe (820) film peak has a ($Q_x$, $Q_y$, $Q_z$) position. Since the azimuthal angle (φ) and sample tilt (χ) are fixed during the scan, the RSM records only the projection of the film peak onto the $Q_x$-$Q_z$ plane, including the effect of any film tilt (a rotation about $Q_y$). To extract accurate lattice constants, the film's reciprocal-space coordinates must be rotated back by the negative tilt angle -α. Because the tilt occurs about $Q_y$, a two-dimensional rotation matrix can be used to perform this correction, and based on the sign convention defined for positive and negative tilt in Figure S1, a clockwise rotation is applied to the measured peak positions:

$$\begin{bmatrix} Q_{x,corrected} \\ Q_{z,corrected} \end{bmatrix} = \begin{bmatrix} \cos(-\alpha) & \sin(-\alpha) \\ -\sin(-\alpha) & \cos(-\alpha) \end{bmatrix} \begin{bmatrix} Q_x \\ Q_z \end{bmatrix}$$

This is equivalent to a counterclockwise 2D rotation matrix with a rotation of α. With these corrected peak positions, one can now calculate lattice constants for the tilted thin film.

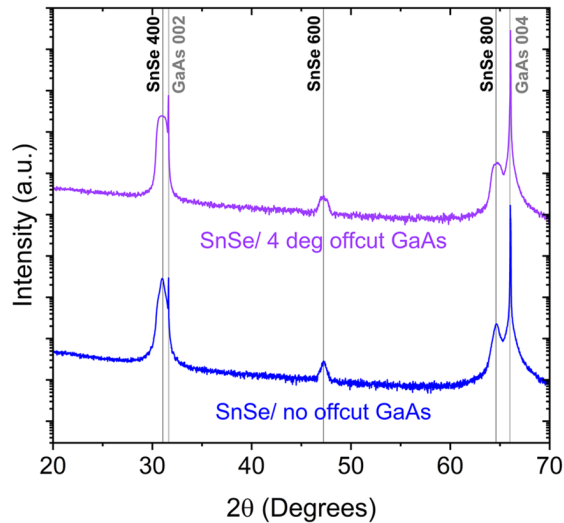

**Figure S2:** symmetric open detector 2θ-ω scans for 25 nm SnSe film on on-axis GaAs(001), and a 70 nm SnSe film on 4 degree offcut GaAs(001). Both have the same OP film orientation with respect to GaAs: only SnSe {100} planes parallel to GaAs{001} planes.

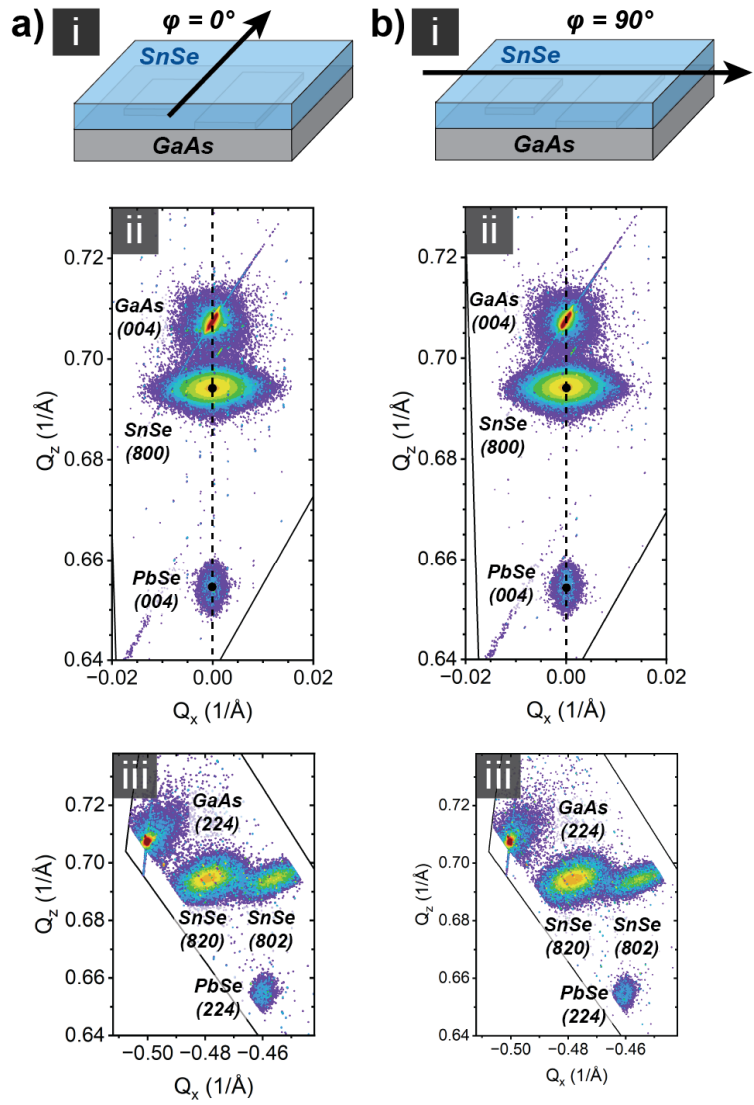

**Figure S3:** RSM of SnSe(~300 nm)/PbSe(15 nm)/GaAs film, with equal fractions of each orthorhombic orientation. (a) φ = 0° and (b) φ = 90°. (i) Schematic showing measurement direction (ii) symmetric RSM and (iii) asymmetric RSM, with film peaks marked.

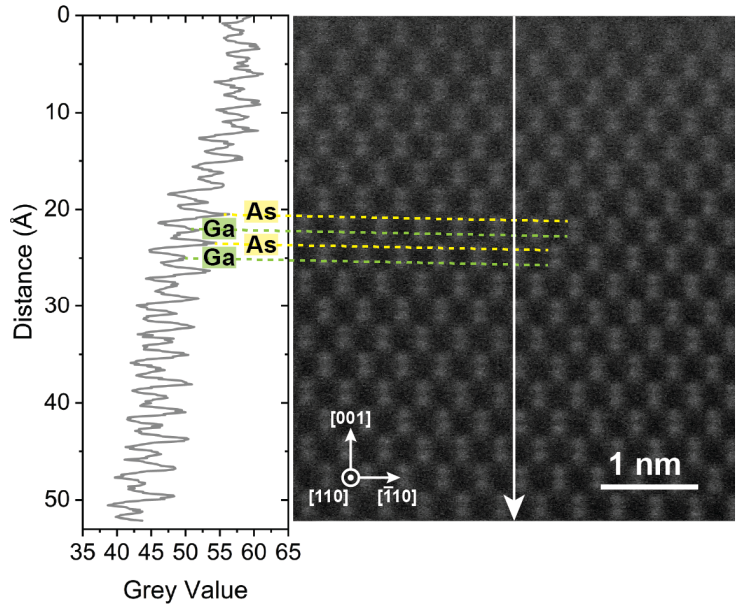

**Figure S4:** Vertical line scan of HAADF intensity of GaAs, averaged over the TEM image width on the right. The periodic intensity modulation reflects the Ga–As dumbbell structure, with the brighter (higher-Z) peaks corresponding to As atoms above Ga atoms.

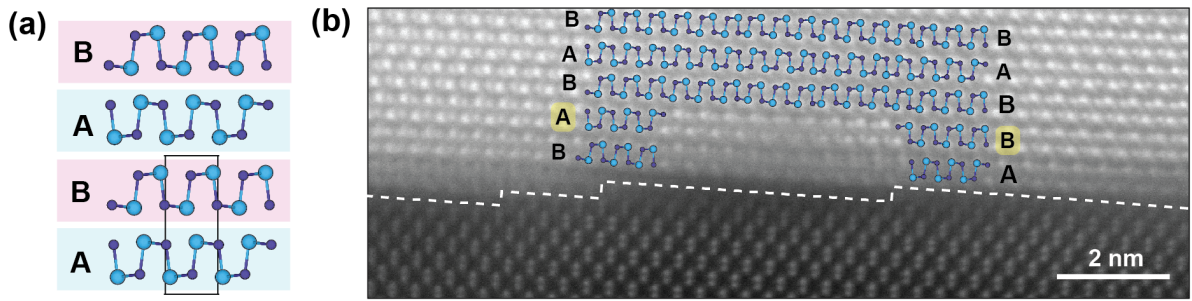

**Figure S5:** Step edges in SnSe grown on 4º offcut GaAs(001). (a) Schematic of alternating "A" and "B" layers. (b) Examples of a SnSe bilayer changing orientation.

5